# Business Rules in e-Government Applications

**Flavio Corradini, Alberto Polzonetti and Oliviero Riganelli**
**Math and Computer Science Department, Università degli Studi di Camerino, Italy**
flavio.corradini@unicam.it
alberto.polzonetti@unicam.it
oliviero.riganelli@unicam.it

**Abstract:** The introduction of Information and Communication Technologies (ICT) into public administrations has been radically changing the way organizations cooperate and, more generally, the way to think about business processes over organizational boundaries. In this paper we describe our approach to combining business processes with business rules in order to integrate effectively single units in an inter- or intra-organizational cooperation. Business rules represent the knowledge that an administration has about its business; with regard to this, they can express strategies, contracts and can influence not only staff relations, but, finally, citizen relations, as well. In other words, business rules are the core of an administration and affect either the business processes or the behaviours of the system participants. They are typically expressed implicitly in business contracts and they are embedded within the source code of many application modules. So a concise and declarative statement of business behaviour is converted into a set of programming instructions, which are spread widely throughout the whole information system. In this way, business rules are difficult to change and keep consistent over the time. For this reason, it is necessary to reengineer the system in order to logically and perhaps physically externalize rules from the application code. In our proposed approach, we describe a cooperation as a collection of tasks combined in certain ways according to the organization logic specified by business rules. Our rule-driven methodology has the goal to make the business process design more adaptable to the changes of internal or external environment.

## 1. Introduction

External organizations, internal policies and law changes put constraints on Public Administrations, exposing them to intensive changes. It is fundamental to have a dynamic system that responds in short time and with low costs of adaptation to these changes. This is possible only if the system is reengineered in a manner in which changes affect only the business rules rather than the whole system: this entails that the development of the PAs towards a digitalized administration requires a deep analysis of its current business processes.

The reorganization of information systems involves a parallel reorganization of both administrative and economic processes with constraints concerning ``legacy'' architectures, organizing structures and current laws. In the last years, business rules are more and more used for the design of information systems, because they are able to make processes flexible and adaptable to changes of external or internal environment. Business rules constraining business processes drive the organizations to the creation and the development of a certain product, so as to achieve predefined business goals with the collaboration of all system participants.

Business rules are typically expressed implicitly in business contracts and they are embedded within the source code of many application modules. So a concise and declarative statement of business behavior is converted into a set of programming instructions, which are spread widely throughout the whole information system. In this way, business rules are difficult to change and keep consistent over the time. For this reason, it is necessary to reengineer the system in order to externalize logically and perhaps physically rules from the application code. In this way, changes may occur directly on the business rules rather on the whole system.

In this paper we describe a software prototype that supports the semi-automatic generation of business processes by the specification of macro activities and suitable business rules. These latter constrain the activities execution such as the temporal order.

## 2. Background

Before defining business rules features, we should briefly define the business process that is the basis of a business purpose planning for an organization. A business process is generally defined as a structured, measured set of activities designed to produce a specified output for a particular customer or market. It implies a strong emphasis on how the work is done within an organization, in contrast to a product focus' emphasis on what work is done (Davenport, 1993).

Generally, besides organizing an administration in single specialized functionalities (such as accounting) and observing the work done by each of them it is important to look at the processes in their completeness. In this way, we could create worthwhile results for citizens. If we are able to manage rules changing quickly, more suitable products for citizen needs can be available (Hammer and Champy, 1993).

Business rules are defined by The Business Rules Group as a statement that defines or constrains some aspect of the business. It is intended to assert business structure or to control or influence the behaviour of the business. The business rules which concern the project are atomic, that is, they cannot be broken down further (The Business Rules Group, 2000). This is their main feature, and for this reason they cannot be broken or divided into details, otherwise important information about business's concepts could be lost. Moreover, they are used as a guide system which influences either the IS perspective or people's behaviours within an organization (business perspective). In fact, business rules can be seen from two different perspectives. The first one is an IS perspective, in which events are registered as data and restrictions about changes in the value of a particular event. The second one is a business perspective, in which rules are bound to stakeholder behaviours in the administration.

Business rules are quite important within an organization for several reasons:
- they can differentiate organizations reducing competition: many organizations actually create a peculiar business rule set, permitting a clear distinction among them;
- they can be reused: they have to be comprehensible and accessible for anybody who wants to manage them. It is unthinkable to use wrong or inconsistent rules only because they are held in the business logic core;
- they adjust to the system in which they are contained and they quickly respond to its changes;
- they allow business decisions to be made in real time.

Business rules can also be internal or external to the Enterprise Architecture (Bajec and Krisper 2005). They are internal when they are defined within the organization and they often derive from strategic elements that show the reason of their existence; they are external when they derive from the external world, including governmental regulations or laws that manage a particular external partner behaviours. For these reasons, business rules represent the knowledge that an organization has about its business; with regard to this, they can express marketing strategies, contracts and they can influence not only staff relations, but also final customer relations. Therefore, rules represent the mechanism by which an organization evolves.

It is very important that an administration redefines relations and automates processes so that it can evolve together with technological innovation. It is necessary, therefore, to construct a new system in which rules are separated from the other components of the system. Extracting rules from business logic implies more flexibility and maintenance across the time of the system, in which they are included: we are proceeding in the Business Rules Approach.
Barbara von Halle affirms that a business rules approach is a methodology—and possibly special technology—by which you capture, challenge, publish, automate, and change rules from a strategic business perspective (von Halle, 2002).

The aim of this type of approach is to create a business rule system that is capable of managing rules separately from other aspects of the system. They are memorized in a data repository and, through a user interface, are made available and accessible to whoever might want to use them. In this way, they acquire a value that is fundamental for the organization. In fact, separating the process flow from the execution of the rules is the most important aspect of the Business Rules Approach because it makes it much easier to process and maintain the system. This creates two internal flows: one for rule management and the other for system process execution. The business planners, having at their disposal a whole flow of rules, have the possibility of distinguishing the dependences that involve business entities. As a result, they dedicate themselves fully to the development and planning of the system without rules, delegating their execution to the rules engine.

## 3. Our approach: The BRAIN

The paradigm of Service-Oriented Computing (SOC) (Singh and Huhns, 2005) is gaining more and more consensuses regarding the integration of heterogeneous information systems. As the same systems can be realized in an independent way, they can be interconnected subsequently through appropriate middleware solutions. Different proposals of standardization have been started (Martin and Burstein et al, 2004; OASIS, 2005). The increasing attention towards these categories of standards reflects the basic connections that exist between the Business Process Management (BPM) (Leymann, Roller and Schmidt 2002) and the Service Oriented Architecture (SOA) (Singh and Huhns, 2005). The emergent technologies of BPM rely on SOA as model for the management of resources, particularly for the software resources, describing process steps or capturing the interactions between a process and its services and users. On the other hand, a service can be also used as a point of access to the fruition of business processes, inducing an intrinsic connection between the service model and the process model.

The reference approaches proposed for services composition are divided mainly into two classes: a centralized composition and a distributed one. The centralized composition is based on workflow model and it models the composition as a process centrally coordinated by a mediator. A "hub and spoke" typology has been adopted. A special role is associated to a service: the "process coordinator". Other services can also communicate, either in terms of control or in terms of information sharing, using the mediator but they cannot communicate with each other. The distributed composition, instead, models the composition as a set of all conversations among participants and does not necessarily require the presence of a centralized coordinator. In this case, a "peer to peer" typology has been adopted. The aim is to capture the behaviours of collaborative processes, during which the interactions among participants are observed from a global perspective.

In order to model business processes, in our research we focalize on a centralized composition based on the workflow model. Workflow Management Coalition (WfMC) defines workflow as the automation of a business process, in whole or part, during which documents, information or tasks are passed from one participant to another for action, according to a set of procedural rules (Leymann, Roller and Schmidt 2002). Therefore a workflow can be seen as a set of organized tasks defining both the order and the conditions on which tasks have to be performed and synchronized. For this reason, we propose a rule driven approach through which a temporal order of tasks can be imposed. As a matter of fact, our approach describes workflows as a set of temporal constraints (business rules) that specify the order in which tasks must be performed. The advantage of using such an approach is that the change of a rule performed by the business manager brings a modification of the process logic, driving to an alignment between the business and the process view.

Organizations are dealing with the challenges of improving the efficiency, rendering, and quality of their services. Such challenges are set out to speed up the response times based on the user needs and to reduce the costs that have built up as a result of the organizational changes. Therefore, we have deduced that it is necessary to reengineer the process of service composition, since the business rules evolve quite quickly with respect to the system in which they have been placed.

In order to address this need we have developed an infrastructure named BRAIN (Business Rules for Adaptive Integration) for the distribution of e-Government services in local administrations based on Shared Services Centre (Corradini et al, 2005), whose goal is to obtain a large rendering of the investments supported by an organization, sharing common elements present in the its single units. In this context, one of the most complex tasks is the service composition. The aim is to manage the cooperation among services provided by different local units in order to supply the most sophisticated functions. For this reason, we need to have shared knowledge to understand the relations among administrations and this can be obtained using a shared business rule repository. We use a specific language WS-BPEL (OASIS, 2005) based on the workflow model to describe composition.

The BRAIN prototype combines business processes with business rules to effectively integrate single units in an inter- or intra- organizational cooperation. In BRAIN, a cooperation is described as a collection of tasks combined in certain ways according to the organization logic specified by business rules. Tasks can be assigned to different service providers and executed in a parallel and distributed environment. The rule implementation takes place through the rules engine, that is a software specialized in rule management and execution. It is also able to make the final product both dynamic and efficient.

## 4. The BRAIN development process

The problem that we have encountered in the deployment of the large amount of existing or new business processes is the lack of a uniform understanding about business and technical requirements, and the relationships among different organizations. In a development process, Public Administration must be aware of its own organizational knowledge as well as its own goals. The development of a coherent view of Public Administration provides a clear understanding of organization and supports the alignment between business and information system perspectives.

Workflow models present a simplified procedural view of the business process. They do not reflect the problematic, contingent nature of organizational work. Many unsuccessful Service-oriented development projects suggest that existing analysis and design methodologies and techniques only cover a part of what is required to support the implementation and deployment of service oriented applications. In this section we want to give a brief description of the development phases of business processes in terms of goals and rules.

## 4.1 Requirements phase

The CHAOS 2000 Survey of the Standish Group reveals that inappropriate requirements or the absence of requirements engineering is one of the main reasons for the unsuccessful software development projects. The primary measure of software development success is the degree to which it meets the purpose for which it was planned.

Goal-directed approaches focus on why systems are designed, expressing the rationale and the motivation to justify software requirements. In this phase, we use scenarios-based design methods, with the important innovation on focusing the scenarios not on tasks in the abstract, but first of all on meeting the goals and needs of specific use cases. Scenarios are behavioral descriptions of a concrete system and its environment arising from restricted situations. They exemplify behaviors enabling hidden needs to be uncovered and are useful for evaluating design alternatives and validating design. In literature, many proposals have been made to couple goals and scenario together. The evident reason is that scenarios and goals have complementary features. Scenarios are concrete, narrative, procedural and make intended implicit properties. Goals are abstract, declarative, and leave intended explicit properties. Scenarios and goals thus complement each other nicely for requirements elicitation and validation. The production of this phase is a requirements definition that aligns the business strategy with the system requirements.

## 4.2 Modelling phase

A Public Administration model is a representation of the organization knowledge about itself or what it would like to become. The development of a coherent view of a Public Administration provides a clear understanding of organization and supports the alignment between business and information system.

The main objectives of business modeling within business process development are threefold:

1. it improves and documents the knowledge regarding the current and the desirable situation of the Business;
2. it reaches a clear and structured set of documents on goals and concepts of the future business;
3. it develops a basis for designing an adequate information system to reach business goals.

In order to extract rules and develop a rule-driven process design, it is necessary to perform a modeling phase that allows the description of abstract or concrete elements of a public administration in a structured and a formal way. In this phase, we focus on the development of models at different layers of abstraction allowing a careful analysis of business rules that govern the processes. In our approach we have recognized the following models:

- **Business Goal Model** aims at describing explicitly in a hierarchical manner the underlying goals of the business. A goal model describes a problem as a composition hierarchy of goals and sub-goals and implies many conditions, for example in a sequence a goal has the condition that the previous goal has been fulfilled;
- **Business Process Model** aims at providing a full description of process in terms of activities and information flow and determines its overall structure (temporal order of activities, cross references among information). The model describes the processes that allow the achievement of public administration goals or sub-goals;
- **Actor Model** consists of identifying, analyzing and describing the actors of processes. It describes how different actors are related to each other and how they are related to goals and processes;
- **Service Model** describes the interaction behavior of a service. It provides the capacity to seize the requirements and the logic base facilitating the reuse of services into the organization;
- **Message Model** describes the content, structure and constraints of information exchanged among services.

**Table 1:** Model formalization

| | Concepts | What | Tools | Languages |
|---|---|---|---|---|
| Goal | Objectives | Scenarios and Goals | Use Case Map and GRL notations | GRL RuleML |
| Process | Processes Activities Events | Control patterns, Communication patterns and Generation patterns | Activity Diagrams | BPEL RuleML |
| Actor | Organizational functions User roles Universe of Discourse | Classes, Associations and Constrains | Activity, Use Case and Deployment Diagrams | WSDL-bindings BPEL-partners RuleML |
| Service | Transaction and Communication Acts | Transaction pattern | Sequence and State diagrams Annotations | RuleML WSDL-messages WSDL-PortType WSDL-operations |
| Message | Message continents and Test Case | Linearization style | OCL expressions, BNF grammar annotations | XML Schema XML infoset RuleML |

In order to manage rules separately, it is necessary to individualize them for each abstraction layer because rules are crosscutting and can be deduced by all models. It is also needed to capture and to map the relationship among model to understand "What affects What", so we are able to estimate the impact of changes in the whole system. After the modeling phase we have created a unique and unambiguous view of administration, for this reason we must also formalize models according to common specification languages as shown in Table 1.

### 4.3 Implementation and integration phases

The implementation phase focus on translating the model specification of each service to an implementation. The various service modules of the modeling are coded into a set of business units.

In other words, implementation is the process that takes a model and transforms it into actual code, ensuring conformity that is judged against the modeling phase. The product of this phase is a set of correct and independent services, which will be then composed in order to obtain more complex and more sophisticated functionalities. In the integration phase the various tested services are connected in an appropriate way. In the process of integration, services are combined with each other to a cohesive whole in order to realize the overall functionalities. In our approach, we give emphasis to requirements and modeling phases because many integration problems can be avoid in these phases. For this reason, a careful analysis of system and its components facilitate the integration phase and so tools for an automatic or semi-automatic integration can be developed.

## 5. Brain architecture

In this section, we present the current architecture of the BRAIN prototype to illustrate the key ideas and concepts in our approach. The architectural diagram of the BRAIN system is presented in Fig. 1. There are four distinct components in the BRAIN Designer architecture, namely, Business Process Designer, Rule Engine, Workflow Pattern Manager and Service Broker.

The proposed architecture for business process design is based on the separation of the rules from the rest of the system. The Rule Engine component handles business rules that are captured and stored in a Rule Repository by the business manager. It is the core of our prototype, each component communicates with it in order to find rules expressing pre- or post- conditions, temporal conditions and other constraints that defines the process behavior and structure. Rules are described by Rule Markup Language (RuleML) (Hirtle et al, 2005). We have distinguished three types of rules: Behavior Process Rule is a complete statement that specifies a temporal constraint among the task events. For example, a business rule may express that if the booking is performed, then the payment can follow;

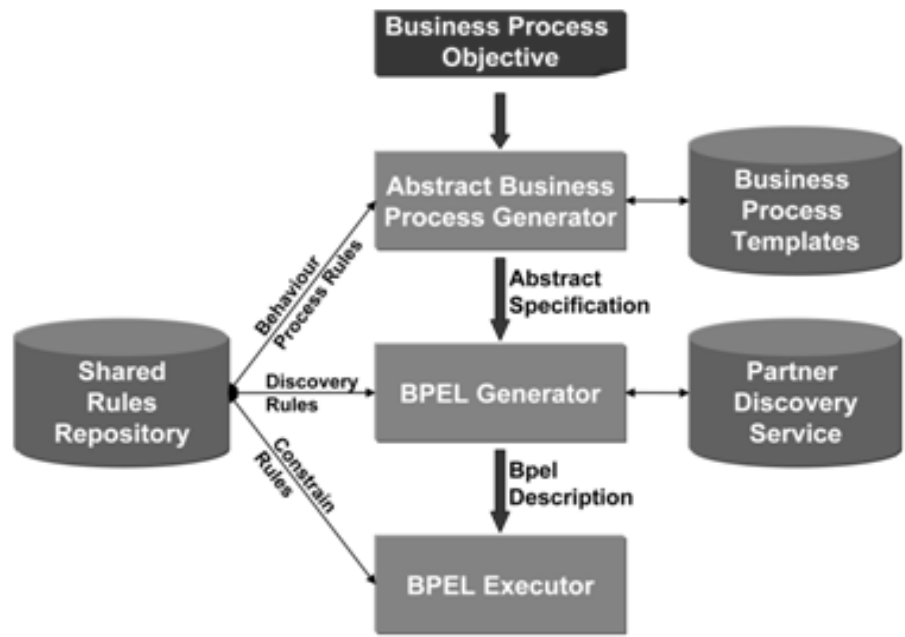

**Figure 1:** Architecture

Constraints Rule is a complete statement that tests conditions and upon finding them true, initiates another business event. For example, if an order is valid then the customer can be notified; Discovery Rule is a complete statement that enables the selection of the suitable service provider. For example, a service provider is appropriate if it can fulfill the order within a period of two hours. The Business Process Designer has the aim to manage the generation process of an executable WS-BPEL. The composition process begins by a request in which the business goals are specified. The designer receives the goals and creates a workflow model derived by the combination of workflow patterns. In this case, the Workflow Pattern Manager selects appropriate patterns stored in the Workflow Patterns Repository according to business goals and related Behavior Process Rules. After this the designer can enrich the model with constrains based on business process needs, in this way it adds decision point to the workflow model generating a WS-BPEL schema. Through the discovery rules, the designer searches partner links that are suitable for the scenario of the pre-defined business process and in this phase the Service Broker supports the designer for the selection. In order to participate in a business process, services must be registered in the Service registry. In this way, the Business Process Designer is able to generate an executable WS-BPEL engine.

## 6. Case study: A tax payment scenario

In order to better understand the BRAIN approach, we show a classic 'Tax Payment' scenario and propose a methodology to reach business goals using "rule-driven" ser-vice composition. Modeling of this process requires many different steps to offer such a full service to citizen. We only use a very simple example with annotations representing the business rules for the different activities, as shown in the Fig. 2. The scenario consists of the following participants:

1. the Citizen orders the payment;
2. the Financial Institution receives the order from a citizen for a payment;
3. the Public Administration is the receiver of the payment;
4. the citizen requires the payment of a tax.

The involved administration receives the request and then sends payment information to citizen. Citizen can now authorize the financial institution to perform the tax payment and the process terminates with the updating of citizen status and the sending of a bill from the hand of administration.

## 7. Demo scenario

In this section we show a prototype for a rule-driven workflow modeling. There are three design steps which produce a WS-BPEL compliant process instance description starting from an abstract workflow model definition.

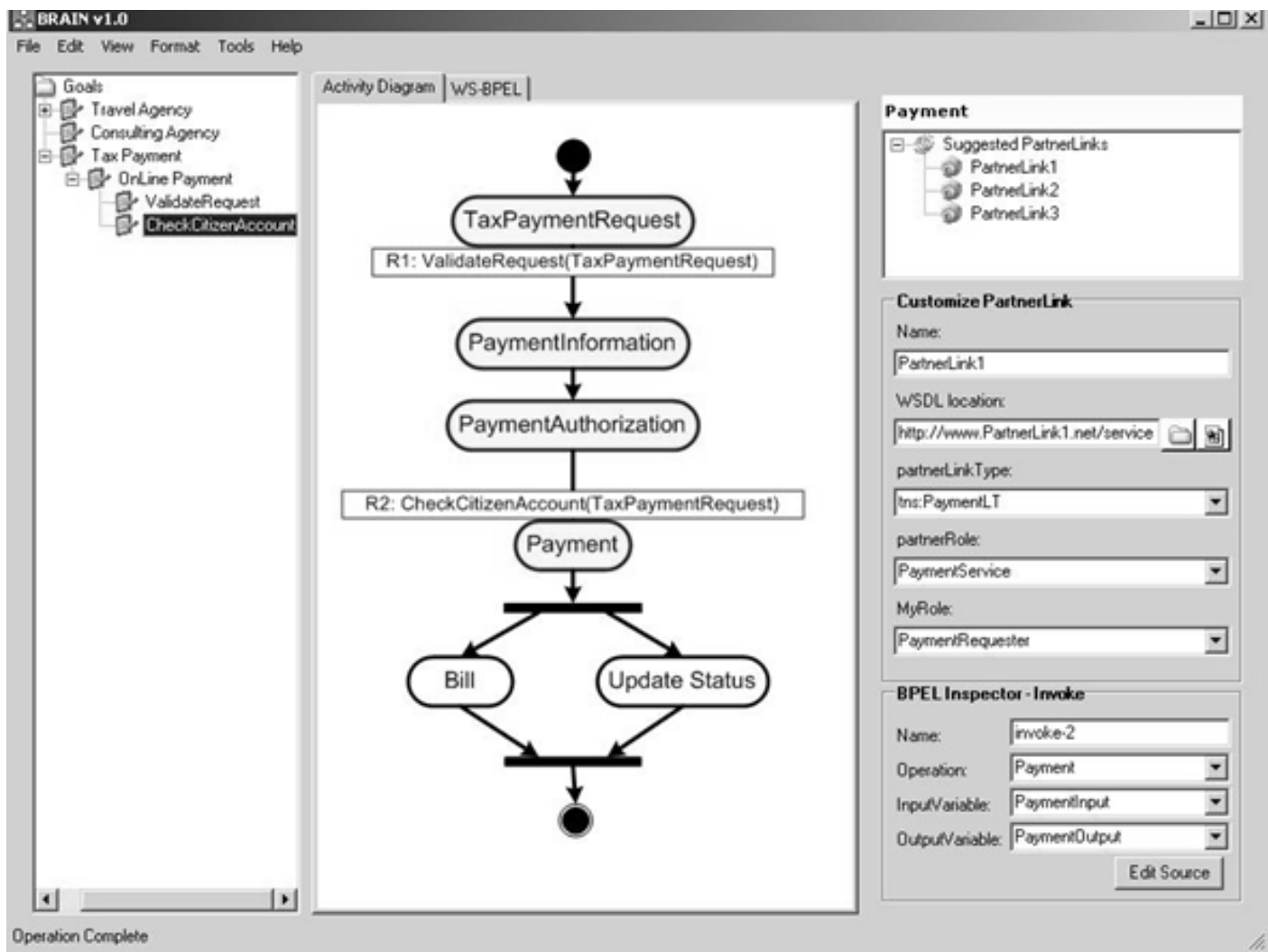

**Figure 2:** BRAIN designer

*Workflow Model Definition*. Figure 2 shows the Business Process Designer layout, a GUI developed for easing the composition procedure: in this first step, we define an abstract workflow model by combining the use of process patterns with the logic of Behavior Process Rules retrieved by the Rule Engine in the Rule Repository. In practice, through a simple selection of goals or sub-goals in the composition area, the activity dependencies are analyzed and shown according to a correspondence among goals, behavior process rules and templates specifications. Finally, we can obtain the complete abstract workflow represented as an activity diagram according to selected business goals. In order to simplify the understanding of process, we use a compact representation of activity diagram.

*WS-BPEL Schema Definition*. This modeling phase concerns the WS-BPEL schema definition. First, a mapping from Activity Diagram specifications to WS-BPEL activities takes place. In particular, we combine Constraints Rules establishing mandatory restrictions or conditions for the behavior of the entire workflow. In this way, we obtain an empty structure of a WS-BPEL document, according to the business process behavior. Finally, we characterize the schema generation procedure by selecting of pre- or post-conditions to be placed among the workflow tasks. In figure 3, we set a post condition R2 for the TaxPaymentRequest activity, which performs a credit check. The R2 rule (CheckCitizenAccount) can be interpreted as a 'reaction rule' in the sense that it can enable an action or not specifying the invocation of actions in response to an event. The action is only performed when a certain condition is satisfied.

*WS-BPEL Process Instance Definition*. For each communication event in the WS-BPEL Model Schema Definition, a set of appropriate service providers (partnerlinks) will be proposed by the service broker, based on discovery rules. For example, in the case study, the payment task can be exploited by invoking a service that can be provided by three possible partner types. Financial institutions are divided into three families according to their service contract. In this way we obtain more dynamicity in the run time phase, because we can change the partnerlink of the same family without needing to redeploy WS-BPEL process. The white activities in figure 3 do not have associated partnerlinks. In this last step, technical specifications of the composition schema can be set (e.g. WS-BPEL activity attributes, input/output variables etc.) through a WS-BPEL document inspector.

## 8. Related works and conclusion

The redesign of government structure and processes is an important topic for the adaptation of IS to the needs of citizens and other stakeholders. In this context, the requirements and modelling phases are very important because many integration problems can be avoided in these phases. For this reason, a careful analysis of a system and its components facilitates the integration making it possible to develop tools for automatic or semiautomatic integration. In this paper we have adopted an experimental methodology in order to redesign the traditional cooperation among information systems. Separating rules from processes allows us to gain better results and makes the IS evolution more understandable and manageable.

Rule-based approaches have been widely accepted as core methodology for modelling and supporting inter- and intra-organizational business processes. In literature several methodologies and techniques are proposed (Casanave, 1995; Ross, 2003; Herbst, 1997). In (Benatallah, Sheng and Dumas, 2003), a peer to peer service composition is used. In this platform ECA rules are used to define the transition among services. In (Herbst and Myrach, 1995) high level model and practice in building rules repositories are described. A contribution of this work is a conceptual schema of business rule relations with the business environment involving processes, organizational units, actors and software components. Many commercial business rule products are also available. One example is ILOG, which is an interesting business rule management and engine allowing the formulation and the implementation of decision services. In our approach, we use business rules to structure and schedule business processes and to describe service selection and service bindings. In particular, we focus on goal oriented methodology to design the process flow. The proposed prototype is a simple goal-oriented tool for the BPEL document generation, with the aim of enabling unskilled managers to rapidly customize their applications to best fit the business process flow of the administration by selecting the required goals. In a public context, the use of a shared repository is important for a global view, so to retain a consistent relationship with the administration environment.

It is obvious that the current standards for service composition are not capable of supporting the complex and dynamic nature of the business process. The combination of business rules with a process-oriented composition allows an adaptive integration of services. A rule-based approach reduces the complexity of such a procedure making it more understandable. Business rules can help to manage Information Systems over time, aligning them with the business of organization.

Though business rules represent the key element of growth and integration in an organization, the rule-based development process of information systems is still not well defined. Actually, information exchange and inter-organizational communication would be difficult without an exhaustive understanding of rules. Future works include the development of an efficient mechanism for the research and the execution of rules. Furthermore, we plan to investigate an effective way to integrate rule engines with WS-BPEL engines in order to increase the flexibility of the whole system not only in the design phase but also in the run-time phase.